\documentclass[pre,amsmath]{revtex4}
\usepackage{graphicx} 
\usepackage{epsfig}

\begin{document}
\title{On the fluid-fluid phase separation in 
charged-stabilized colloidal suspensions
}
%\shorttitle{Phase stability of charged colloidal suspensions}
\author{Yan Levin$^{1,2}$\thanks{E-mail: \email{levin@if.ufrgs.br}}, 
    Emmanuel Trizac$^2$\thanks{E-mail:
    \email{Emmanuel.Trizac@th.u-psud.fr}}, 
Lyd\'eric Bocquet$^3$\thanks{E-mail: \email{Lyderic.Bocquet@lpmcn.univ-lyon1.fr}}}

\affiliation{$^1$ Instituto de F\'{\i}sica, 
  Universidade Federal do Rio Grande do Sul \\ 
  Caixa Postal 15051, CEP 91501-970, Porto Alegre, RS, Brazil}   
        
\affiliation{$^2$ Laboratoire de Physique Th\'eorique, UMR CNRS 8627, Universit\'e
de Paris XI, B\^atiment 210, F-91405, Orsay Cedex, France}

\affiliation{$^3$LPMCN, UMR CNRS 5586,
Universit\'e Claude Bernard Lyon 1, 
43 bld du 11 novembre 1918,
69622 Villeurbanne Cedex, France
}
%\pacs{71.10.C}{Electron gas-theories and models}
%\pacs{64.60.Cn}{Order-isorder transformations; statistical mechanics of
%model systems}

\date{\today}

\begin{abstract}
We develop a thermodynamic description of particles held at a fixed surface
potential.  This system is of particular interest in view of the
continuing controversy over the possibility of a fluid-fluid
phase separation in aqueous colloidal suspensions 
with monovalent counterions.
The condition of fixed surface potential allows in a natural way
to account for the colloidal charge renormalization.  In a first approach,
we assess the importance of the so called ``volume terms'', and find that
in the absence of salt, charge renormalization is sufficient 
to stabilize suspension against a fluid-fluid phase separation.  Presence
of salt, on the other hand, is found to lead to an instability.  
A very strong dependence on the approximations
used, however, puts the reality of this phase transition in a serious
doubt. To further understand the nature of the instability we 
next study a Jellium-like approximation, which does not
lead to a phase separation and produces a  relatively accurate
analytical equation of state for a deionized 
suspensions of highly charged colloidal spheres.
A critical analysis of various theories of strongly asymmetric
electrolytes
is presented to asses their reliability as compared to the 
Monte Carlo simulations.

\end{abstract}

\maketitle

%%%%%%%%%%%%%%%%%%%%%%%%%%%%%%%%%%%%%%%%%%%%%%%%%%%%%%%%%%%%%%%%%
\section{Introduction}

There is a long standing debate in the field of colloidal science
concerning a possibility of phase separation in aqueous colloidal
suspensions containing monovalent counterions 
\cite{vRoij1,vRoij2,Warren,Linse,Levin1,Tamashiro1,Diehl,Grunberg,Deserno,Bhuiyan,BSPB,Tellez,%
Dufreche,Tamashiro}. 
Charged colloids
provide a particular challenge to the theorists.  The long ranged
Coulomb interaction and the extremely large asymmetry between the
polyions and the counterions makes it very difficult to apply to this
system the traditional methods of the liquid state theory.  In
particular, such a well established tool of condensed matter theorists
as integral equations is found to be useless when applied
to strongly asymmetric electrolytes at large couplings (low temperatures).  
For a wide range of  parameters of
physical interest 
the integral equations  fail to even converge. The original
hope of associating the lack of convergence with an underlying physical
instability has proven to be unfounded\cite{Fis81,Bel93}.

Recently, a linearized 
density functional theory has been used to study charged 
colloidal suspensions
\cite{vRoij1,vRoij2}.
The theory predicted a thermodynamic instability which 
manifested itself as a fluid-fluid phase separation.  
However,
the underlying approximation of the theory put its conclusions in
doubt \cite{Diehl,Grunberg,Deserno}.  Specifically the linearization of 
the density functional
lowers the statistical weight of the configurations in which
the counterions are in a close vicinity of the polyions.  This effect
can be partially accounted for through the renormalization of the
bare colloidal charge \cite{Alexander,Belloni}.  Unlike the bare charge, which 
can be very large, the effective (renormalized) charge is 
found to be bounded by the saturation value controlled 
---in a given solvent--- by the colloidal size,
temperature, and salt concentration.  The renormalization
of colloidal charge was argued to wash out the phase transition
predicted by  the linear theories \cite{Diehl}.  
Furthermore,  numerical solutions
of the full non-linear Poisson-Boltzmann (PB) equation inside a Wigner
Seitz (WS) cell shows
absence of any instability\cite{Grunberg,Deserno,Tellez}.  
On the other hand, linearization of the PB equation leads to
a non-convex pressure, as a function of colloidal concentration, 
similar to the
one observed in other linear theories. 
All these suggest that the phase instability predicted by the linear
theories might be an artifact 
of the underlying approximations.

To further explore these interesting points,  
we have investigated
the thermodynamics of particles fixed at 
constant surface potential\cite{Trizac1,Trizac2,Levin}. 
Relevance of the constant potential ensemble
follows from the observation that
unlike the bare charge, the effective charge of 
colloidal particles
does not grow indefinitely but instead saturates.  
The saturation value is such that the effective
electrostatic potential $\phi_s$ in the
vicinity of colloidal surface is $\beta q \phi_s \approx 4$, where
$\beta=1/k_B T$ and $q$ is the elementary charge. 
Thus, although the colloidal charge can be 
very large, the potential near the colloidal surface, i.e. within
the Debye length, does not increase beyond $\beta q \phi_s \approx 4$.  This
simple observation is sufficient to construct a consistent thermodynamic
description of colloids with a state dependent effective charge.
Since the WS cell description of colloidal suspension\cite{Marcus} 
does not lead to a fluid-fluid phase separation (see
\cite{Tellez} for a general argument), to further understand the 
mechanism of the 
instability observed within the linearized
theories, we focus on this alternative treatment.

We shall first (section \ref{sec:theory}), 
study the effect of charge 
renormalization on the polyion-microion interaction free energy, i.e. the 
volume term that appears when the original mixture 
of colloids, coions and counterions
is mapped onto an effective one component system of dressed colloids
\cite{vRoij1,vRoij2,Levin1}. 
%It should be, however, kept in mind that the effective 
%charges are intended to describe the ``far field''
%features of the electrostatic interactions and care must
%be taken when consideration of ``near field'' properties
%are needed.  
In section \ref{sec:jellium} an alternative 
derivation of the thermodynamic equation of state based only on the far 
field considerations is presented. The corresponding pressure-density 
isotherms, in this case, do not exhibit criticality at 
any salt concentration. 
In order to assess the reliability of various  approaches,
in section \ref{sec:discussion} we compare the corresponding pressures
to the results of the Monte Carlo simulations of  Linse \cite{Linse}.
We also consider the, recently proposed, 
symmetric Poisson-Boltzmann \cite{Bhuiyan} and
the ``boot-strap'' Poisson-Boltzmann \cite{BSPB} theories. 
%In light of the existing 
%Monte Carlo data, we conclude that the simple Poisson-Boltzmann (PB) cell
%model is by far the most accurate and reliable, even at low densities
%and up to relatively high electrostatic coupling. 
Conclusions are drawn in section \ref{sec:conclusion}.

%%%%%%%%%%%%%%%%%%%%%%%%%%%%%%%%%%%%%%%%%%%%%%%%%%%%%%%%%%%%%%%%%%%%
\section{Role of volume terms }
\label{sec:theory}

\subsection{State dependent effective charges}

Consider a colloidal suspension at concentration $\rho_p$, 
containing spherical polyions
of charge $-Zq$ and radius $a$ 
in contact with a monovalent salt reservoir at concentration $c_s$.
Now suppose one colloid is fixed at $r=0$. In a continuum
approximation consisting of smearing out the charge of other colloids and 
linearizing the  Poisson-Boltzmann
equation\cite{Levin}, 
the electrostatic potential at distance $r$ from the center
of colloid is
%-----------------------------
\begin{eqnarray}
\label{1}
\phi(r)=-\frac{Zq \theta(\kappa a) e^{-\kappa r}}{\epsilon r} \;, 
\;\;\;\; 
\theta(x)=\frac{e^{x}}{(1+x)}\;,
\end{eqnarray}
%--------------------------------
where the inverse Debye length is
%-----------------------------
\begin{equation}
\label{2}
\kappa = \sqrt{4 \pi \lambda_B (\rho_++\rho_-}) \;,
\end{equation}
%--------------------------------
and the Bjerrum length is 
%-----------------------------
\begin{equation}
\label{5}
\lambda_B=\frac{q^2}{\epsilon k_B T}\;.
\end{equation}
%--------------------------------
The mean densities of coions
and counterions inside the suspension  are respectively 
$\rho_-$ and $\rho_+$.   When $\rho_p \to 0$,
$\kappa^2 \to \kappa_s^2 = 8 \pi \lambda_B \,c_s$. 

For highly charged polyions, Eq. (\ref{1})
strongly overestimates
the real electrostatic potential. However, it can be made consistent
with the full non-linear PB, if instead of the bare charge $Z$ an effective,
renormalized, charge $Z_{eff}$ is used.  The observation that
for large surface potentials the
electrostatics away from the colloidal surface is completely
insensitive to the surface charge density allows for the
``far-field''  definition of the effective charge. Specifically,
viewed from a distance larger than the Debye length, and provided
that $\kappa a>1$,
the surface potential of a strongly charged colloidal particle
appears to be $\beta q \phi(a) \approx -4$ \cite{Trizac1,Trizac2,Levin}. 
Combining this with 
Eq. (\ref{1}) leads directly to \cite{Trizac1,Levin}
%-----------------------------
\begin{equation}
\label{3}
Z_{eff}=\frac{4 a }{\lambda_B}\,(1+\kappa a)\;.
\end{equation}
%--------------------------------
We should note that this is the saturated value of the 
effective charge relevant for the highly
charged colloidal particles.  
For weekly charged particles there is little or no
charge renormalization. In the infinite dilution
limit\cite{JPA} of one colloid immersed in a 1:1 electrolyte of concentration
$c_s$, the exact result for the saturation limit of $Z_{eff}$ is
\begin{equation}
Z_{eff} = \frac{4 a }{\lambda_B} \left(\frac{3}{2} + \kappa_{s} a \right).
\label{eq:Zeff}
\end{equation}
This expression --valid up to corrections of order $(\kappa a)^{-1}$,
which turn out to be quite small  as soon as $\kappa a>1$, is very close
to the approximation (\ref{3}), derived from the matching procedure 
detailed in \cite{Trizac2,Levin}.

The effective charge depends strongly on the electrolyte concentration
inside the suspension.  Salt screens the electrostatic interactions
between the counterions and the polyions and leads to an increase in the
colloidal effective charge.  While the effective charge of colloidal
particles is a strongly state dependent function, the effective 
surface potential is not.  Therefore, inside  suspension, colloids 
behave as if their surface potential was effectively fixed.  It is interesting,
therefore, to study the thermodynamics of colloidal particles at
fixed surface potential.

%%%%%%%%%%%%%%%%%%%%%%%%%%%%%%%%%%%%%%%%%%%%%%%%%%
\subsection{Thermodynamics of particles at fixed surface potential}

The change of the thermodynamic ensemble from the constant charge
to the constant surface potential allows us in a natural way to
explore the role of charge renormalization in the framework of a linear
theory. The surface potential is related to the effective colloidal charge
through the Eq. (\ref{1})
%-----------------------------
\begin{equation}
\label{4}
\varphi=-\beta q \phi(a)=\frac{Z \lambda_B}{a (1+\kappa a) }\;.
\end{equation}
%--------------------------------

As mentioned above, the reduced surface potential
within the non-linear Poisson-Boltzmann theory 
is found to saturate at $\varphi=4$.  For the sake of generality
we shall, however, keep its value arbitrary. In the subsequent 
analysis, $Z$ will refer to the saturation value of the colloidal 
effective charge. An implicit assumption is therefore that 
the bare charge largely exceeds the effective one. 

In the simplest approximation, the 
Helmholtz free energy of the suspension is a sum of entropic and
electrostatic contributions
%-----------------------------
\begin{equation}
\label{6}
\beta F=N_+[\ln{\rho_+ \Lambda^3}-1]+
N_-[\ln{\rho_- \Lambda^3}-1]+\beta F^{el}\;,
\end{equation}
%--------------------------------
where $\Lambda$ is the de Broglie thermal wavelength, and
$N_+$ and $N_-$ refer to the number of 
counterions and coions inside the suspension.

The electrostatic free energy results from the polyion-microion,
microion-microion 
and  the polyion-polyion interactions.  For suspensions containing
monovalent counterions the polyion-counterion interaction is the 
dominant contribution 
and will be the only one kept in the present exposition (we come back to this 
point in section \ref{sec:discussion}).  We find\cite{Levin}
%-----------------------------
\begin{equation}
\label{7}
\beta F^{el}=\frac{Z^2 \lambda_B N_p}{2 a (1+\kappa a)}\;.
\end{equation}
%--------------------------------  
This expression can be obtained through the usual Debye charging
process in which all the particles are simultaneously charged
from $0$ to the their final charge\cite{Levin}. Alternatively
a surface charging process, at constant Debye length can
be employed\cite{PRE97}. It is noteworthy that in the
salt free case, we recover precisely the volume term 
obtained in Refs \cite{vRoij1,vRoij2}. 

If the suspension is in contact with a salt reservoir
of chemical potential $\mu_s$,
the effective charge of colloidal particles, as well as the
number of counterions and coions, is determined by the minimum
of the grand potential function
%-----------------------------
\begin{equation}
\label{8}
\beta \Omega=\beta F-\beta\mu_s (N_++N_-)-(\varphi+\gamma) Z N_p
-\omega N_p\left[\varphi(1+\kappa a)-\frac{Z \lambda _B}{a}\right]-
\gamma(N_+-N_--ZN_p)\;.
\end{equation}
%--------------------------------  
In this equation, $\omega$ and $\gamma$ are the Lagrange multipliers: 
$\gamma$ ensures the charge neutrality 
of the system, while $\omega$ enforces the relationship between
the surface potential and the effective charge, Eq. (\ref{4}).
In the biophysics literature  
$\gamma$ is known as the Donnan potential. It results from the 
inability of macroions to diffuse through a semi-permeable membrane.
This is precisely the situation that we have in mind, while the
microions are assumed to be in a free exchange with the reservoir,
the polyions are confined to the interior of suspension.  This 
restriction on the polyion mobility 
results in a potential difference $\gamma$, 
between the bulk of suspension
and the reservoir. The colloids are then held at potential 
$\varphi+\gamma$ with respect to the reservoir, or equivalently 
at potential 
$\varphi$ with respect to the bulk of the suspension. Within the
WS cell model, a similar prescription of constraining
the potential difference between the colloidal surface and
the outer (reservoir) boundary has been shown to yield a surprisingly
good agreement with the full non-linear 
Poisson-Boltzmann equation\cite{Trizac2}.

Minimizing the grand potential with respect to $N_+$, $N_-$ and $Z$, we find
%-----------------------------
\begin{equation}
\label{9}
\frac{\partial \beta\Omega}{\partial N_\pm}=\frac{\partial \beta F}{\partial
N_\pm}-\mu_s-
\omega \left[\frac{2 \pi \varphi \lambda_B a \rho_p}{\kappa} \right] \mp\gamma =0\;,
\end{equation}
%--------------------------------  

%-----------------------------
\begin{equation}
\label{11}
\frac{\partial \beta\Omega}{\partial Z}=\frac{\partial \beta F}{\partial
Z}-\omega N_p \left[\frac{\lambda_B}{a} \right]-\varphi N_p =0\;,
\end{equation}
%--------------------------------
and the charge neutrality condition reads
%-----------------------------
\begin{equation}
\label{12}
N_+-N_-=\frac{\varphi a (1+\kappa a)}{\lambda_B} N_p\;.
\end{equation}
%--------------------------------
Noting that 
%-----------------------------
\begin{equation}
\label{13}
\frac{\partial \beta F}{\partial Z}=\varphi N_p\;,
\end{equation}
%--------------------------------                
Eq. (\ref{11}) simplifies to
%-----------------------------
\begin{equation}
\label{14}
\omega=0\;.
\end{equation}
%-------------------------------- 
Eliminating the Lagrange multiplier between Eqs. (\ref{9}), 
we are left with two equations which govern the 
concentrations of counterions and coions inside suspension,
%-----------------------------
\begin{equation}
\label{15}
\rho_+ \rho_-=c_s^2 \exp\left(\frac{2 \pi \varphi^2 a^2 \rho_p}{\kappa}\right)\;
\end{equation}
%--------------------------------
and
%-----------------------------
\begin{equation}
\label{16}
\rho_+- \rho_-=\frac{\varphi a (1+\kappa a)}{\lambda_B} \rho_p \;.
\end{equation}
%--------------------------------   

%%%%%%%%%%%%%%%%%%%%%%%%%%%%%%%%
\subsection{The equation of state}
      
The osmotic pressure inside suspension of
colloids at fixed surface potential is
%-----------------------------
\begin{equation}
\label{17}
P=-\frac{d \Omega}{d V}
\Big\arrowvert_{N_p,\mu_s,\omega,\gamma,\varphi}.
\end{equation}
%-------------------------------     
It is important to keep in mind 
that as  $\Omega$ changes with volume, the
number of coions, counterions, as well as the charge of colloidal
particles 
are all varying. This is the reason for writing the total derivative
in the expression (\ref{17})
%-----------------------------
\begin{equation}
\label{18}
\frac{d \Omega}{d V}
\Big\arrowvert_{N_p,\mu_s,\omega,\gamma,\varphi}=
\frac{\partial \Omega}{\partial N_+}\frac{d N_+}{d V}+
\frac{\partial \Omega}{\partial N_-}\frac{d N_-}{d V}+
\frac{\partial \Omega}{\partial Z}\frac{d Z}{d V}+
\frac{\partial \Omega}{\partial V}.
\end{equation}
%-------------------------------
Recalling that at the thermodynamic equilibrium
%-----------------------------
\begin{equation}
\label{19}
\frac{\partial \Omega}{\partial N_+}=
\frac{\partial \Omega}{\partial N_-}=
\frac{\partial \Omega}{\partial Z}=0
\end{equation}
%-------------------------------
expression for pressure simplifies to 
%-----------------------------
\begin{equation}
\label{20}
P=-\frac{\partial \Omega}{\partial V}
\Big\arrowvert_{N_p,N_+,N_-,Z,\mu_s,\omega,\gamma,\varphi}=
-\frac{\partial F}{\partial V}\Big\arrowvert_{N_p,N_+,N_-,Z}.
\end{equation}
%-------------------------------
Equation (\ref{20}) beautifully illustrates the thermodynamic principle of 
ensemble equivalence.  The functional form of the pressure 
is the same weather the calculation is done in the fixed potential 
ensemble using the grand potential function $\Omega$, or in the
fixed colloidal charge ensemble using the Helmholtz free energy $F$.
We stress that simply inserting $Z_{eff}(V)$ with its state
dependence into  $F$ and then differentiating
it with respect to volume will lead to an incorrect result.
If the Helmholtz free energy is used, the variation must be performed
at {\it fixed} colloidal charge.

In general it can be very difficult to find a suitable thermodynamic 
potential for a constrained system.  The calculation of pressure,
on the other hand, can be done very straightforwardly using the constant $Z$
ensemble, and enforcing the constraint {\it a posteriori}.
Evaluating the partial derivative in Eq.~(\ref{20}), the osmotic
pressure inside the suspension takes a particularly simple 
form
%-----------------------------
\begin{equation}
\label{21}
\beta P= \rho_++\rho_--\frac{1}{4} \frac{a}{\lambda_B} \varphi^2 \kappa
a\rho_p \;,
\end{equation}
%-------------------------------
where the concentrations of coions and counterions 
are determined from Eqs. (\ref{15}) and (\ref{16}).

\begin{figure}[htbp]
\includegraphics[width=8cm]{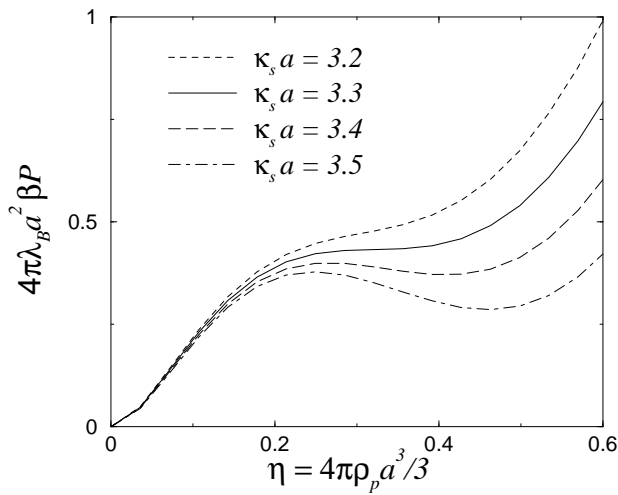}
\caption{Pressure isotherms for different reservoir salt concentrations 
($\kappa_s^2 = 8 \pi \lambda_B c_s$, $\varphi=4$)}
\label{fig:1}
\end{figure}

\begin{figure}[htbp]
\includegraphics[width=8cm]{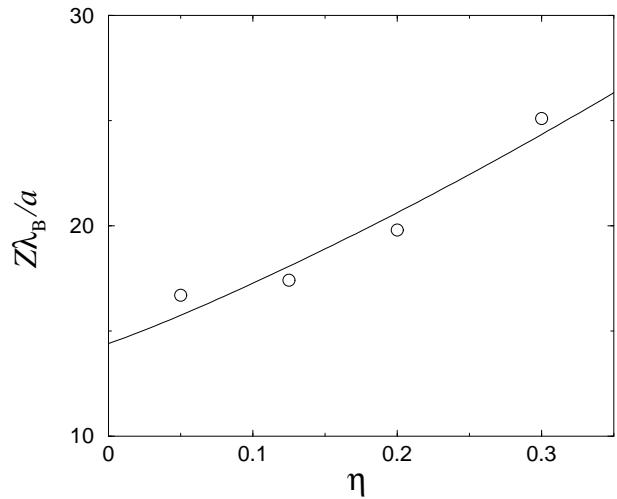}
\caption{Comparison between the effective charge (solid line) 
found using our variational approach,
to that obtained within the Poisson-Boltzmann 
cell theory (circles), following the prescription
proposed by Alexander {\it et al.} \cite{Alexander,Langmuir}}
\label{fig:10}
\end{figure}

In the special case of vanishing salt concentration, 
Eqs. (\ref{15}) and (\ref{16}) simplify  to
%-----------------------------
\begin{equation}
\label{22}
\rho_-=0 \;,
\end{equation}
%------------------------------- 
%-----------------------------
\begin{equation}
\label{23}
\rho_+=\varphi \rho_p\frac{a}{\lambda_B}\left[1+\frac{3}{2} \eta \varphi
+
\frac{1}{2}\sqrt{3 \eta \varphi(4+3 \eta \varphi)}\right]\;,
\end{equation}
%------------------------------- 
and the ratio of colloidal size to Debye length is
%-----------------------------
\begin{equation}
\label{24}
\kappa a = \frac{3 \eta \varphi}{2} + 
\frac{1}{2}\sqrt{3 \eta \varphi(4+3 \eta \varphi)}\;,
\end{equation}
%-------------------------------
where $\eta = 4 \pi \rho_p a^3/ 3$ is the macroion volume fraction.
For salt-free suspensions pressure becomes
\begin{equation}
4 \pi \lambda_B a^2 \beta P = 3 \eta \varphi 
\left(1+\kappa a -\frac{1}{4} \varphi \kappa a \;.
\right)
\label{eq:pressZ}
\end{equation}
%or, in terms of rescaled charge $\widetilde Z  = Z \lambda_B/a$
%\begin{equation}
%4 \pi \lambda_B a^2 \beta P = 3 \eta \widetilde Z 
%\left[
%1-\frac{1}{4} \frac{\widetilde Z 
%\sqrt{3 \eta \widetilde Z}}{\left(1+\sqrt{3 \eta \widetilde
%Z}\right)^2}
%\right].
%\end{equation}
If $\varphi < 4$ the osmotic pressure
is a convex up function of colloidal density. For  
real colloids with $\varphi=4$, the
pressure is a linear function of colloidal 
density, $\beta P = 4 \rho_p a/\lambda_B$.
For surface potentials strictly
above $4$, a thermodynamic instability appears.  
It is very curious to note
that the instability sets in precisely at $\varphi=4$, which is the
saturation value for the surface potential obtained within the
non-linear Poisson-Boltzmann theory. 
Renormalization of the electrostatic free energy is, therefore, sufficient
to stabilize a salt free {\it real} colloidal suspension ($\varphi=4$) 
against a fluid-fluid
phase separation\cite{Diehl}.
However, the fact that the critical surface potential is precisely equal
to the saturation value of the non-linear theory, suggests that
the approach, most likely, is very sensitive to the approximations made.  
Furthermore, a relatively small amount of salt 
destabilizes suspension even when $\varphi=4$, see Fig. \ref{fig:1}. 
The critical salt concentration 
for highly charged colloids of radius $a=1000\,$\AA, is $c_s^*\approx
10^{-4}$ M, which corresponds
to $\kappa_s^* a \simeq 3.3$.
Since there is no explicit  polyion-polyion 
nor  microion-microion interaction, 
the instability is completely driven by the
polyion-counterion correlations. 
%We also observe that for
%$\kappa_s^* a <3.3$, a slight increase from $\varphi=4$ to 
%$\varphi > 4$ destabilizes the suspension, as was the case without salt.
In Fig. \ref{fig:10} we show the effective charge
$Z$ resulting from our approach, as a function of volume fraction.
A good agreement with the Poisson-Boltzmann cell model
is found. 

\begin{figure}
\includegraphics[width=8cm]{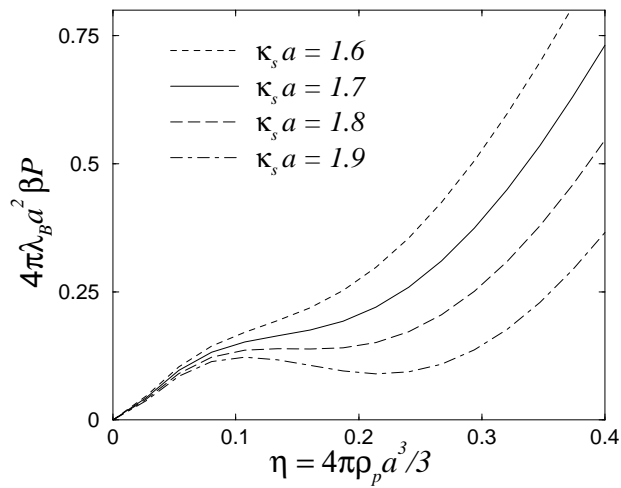}
\caption{Pressure isotherms for different reservoir salt concentrations,
making use of Eq. (\ref{eq:Zeff}) instead of Eq. (\ref{3}). 
The situation is now
different from that of the previous constant $\varphi$ ensemble. Working is the
constant $Z$ ensemble with the Helmholtz free energy (\ref{6}) nevertheless
allows to compute the pressure. }
\label{fig:2}
\end{figure}

It is important to stress that even a minor modification of 
the approximations employed
may have a dramatic effect on the predicted phase instability. 
One may wish for instance to use Eq. (\ref{eq:Zeff})
for the effective charge instead of Eq. (\ref{3}).
It is important to note, however, that such a 
modification means that colloid is no longer held at fixed  
potential. Therefore,   
the grand potential $\Omega$, as written in (\ref{8}), 
can no longer be used.
However,
we can compute the
functional dependence of $P$ on $Z$, $\rho_p$ and $c_s$ by differentiating
the Helmholtz free energy $F$ with respect to
volume at {\it constant}  $Z$, and enforcing the constraint
$Z = a (4 \kappa a + 6)/\lambda_B$ {\it a posteriori}. Following this route,
we recover the same equation of state as before [i.e. Eq. (\ref{eq:pressZ}) 
in the salt free case but with now a different salt dependence of $Z$] and
the  critical salt concentration above which the instability sets in (see
Figure \ref{fig:2}), decreases by a factor of four to  
$\kappa_s^* a \simeq 1.77$.

%%%%%%%%%%%%%%%%%%%%%%%%%%%%%%%%%%%%%%%%%%%%%%%%%%%%%%%%%%%%%%%%%%%%%%%%
\section{An alternative approach : a jellium approximation}
\label{sec:jellium}

In the previous section, we have 
found a fairly accurate expression for the effective colloidal 
charge at saturation and used it to renormalize
the electrostatic free energy. 
It is important to remember, however, that the effective charge,
is by definition related to the  ``far field'' asymptotic
properties of the electrostatic potential. Its use  
for the renormalization of electrostatic free energy is, therefore,
questionable, since not only far field but also the 
near field properties of the electrostatic potential may  be relevant.
We now turn our attention to a simple approach which 
relies only on
the far field features of the electrostatic potential 
to obtain the equation of state. As within the WS
cell picture, use will be made of contact theorem, which
relates the osmotic pressure to the concentration of counterions
in the region where the electric field is zero\cite{Marcus}.

We now reconsider the approach put forward at the beginning of section
\ref{sec:theory}.  Consider one colloidal particle fixed
at the origin of coordinate system. As before the charge of other
microions is uniformly smeared 
throughout the solution. On the other hand, positions of counterions are
strongly correlated with those of  colloids. 
The system then forms a jellium, where the electrostatic potential
far from the colloid $\phi_\infty$ (bulk) differs from that
in the reservoir (chosen to vanish).
The solution of the linearized Poisson-Boltzmann equation 
for $\delta \phi = \phi-\phi_\infty$ is again given
by Eq. (\ref{1}), where the screening length is related 
to the bulk salt concentration, 
$\kappa^2 = 4 \pi \lambda_B (\rho_+(\infty)+\rho_-(\infty))$.
In the spirit of the previous discussion, the colloidal particles
are held at constant surface potential $\varphi=4$ with respect 
to the bulk $\phi_\infty$, 
which again imposes
$Z = 4 a (1+\kappa a)/\lambda_B$. 
The counterions and coions are distributed inside the
jellium in accordance with the Boltzmann distribution
%-----------------------------
\begin{equation}
\label{24a}
\rho_+(r)=c_s e^{-\beta q \phi(r)} 
\end{equation}
%-------------------------------

%-----------------------------
\begin{equation}
\label{24b}
\rho_-(r)=c_s e^{+\beta q \phi(r)} 
\end{equation}
%-------------------------------
where $\phi(r)$ is the local electrostatic potential with respect
to the reservoir. Taking the product of Eqs.(\ref{24a}) and (\ref{24b}) 
we find the familiar condition for Donnan equilibrium
$\rho_+(r) \rho_-(r) = c_s^2$. The electro neutrality constraint
$\rho_+(\infty) = \rho_-(\infty)  + Z \rho_p$, 
closes the problem\cite{JPCM}.
The total concentration of microions inside the suspension
is then related to their concentration inside the salt
reservoir through
\begin{equation}
(\kappa a)^4 = (\kappa_s a)^4 + [12 \eta (1+ \kappa a)]^2.
\label{eq:kappadonnan}
\end{equation}
The osmotic pressure, within the non-linear PB theory,
is determined from the concentration of microions in the region
where the electric field is zero.  Since the electrostatic
potential decays exponentially with $r$, 
the electric field vanishes when $r\to \infty$.
Within  the jellium approximation the osmotic
pressure then takes a particularly simple form,
%-----------------------------
\begin{equation}
\label{24c}
\beta P = \rho_+(\infty) + \rho_-(\infty)=\frac{\kappa^2}{4 \pi \lambda_B}. 
\end{equation}
%-------------------------------
One can show that solution of Eq. (\ref{eq:kappadonnan}) 
obeys the inequality
\begin{equation}
\frac{\partial \kappa}{\partial \eta}\biggl|_{\kappa_s} >0
\end{equation}
which ensures that the compressibility is always positive
and that suspension is stable against the phase separation. 
In the absence of salt, we obtain a simple analytic expression
\begin{equation}
4 \pi \lambda_B a^2 \beta P \,=\, 12 \eta \left[
1+6 \eta + \sqrt{12 \eta (1+3 \eta)}
\right],
\label{eq:pressjel}
\end{equation}
that will be tested against experimental data in section \ref{sec:discussion}.

%%%%%%%%%%%%%%%%%%%%%%%%%%%%%%%%%%%%%%%%%%%%%%%%%%%%%%%%%%%%%%%%%%%%%%%%
\section{Discussion}
\label{sec:discussion}

%%%%%%%%%%%%%%%%%%%%%%%%%
\subsection{Comparison with the Monte Carlo simulations}

\begin{table}
\begin{center}
\begin{tabular}{|c|c|c|c|c|}
\hline
$~~~~\lambda_B/a$~~~~ & ~~~~$\beta P/n$,~MC~~~~&
 ~~$\beta P/n$,~$Z=Z_{\hbox{\scriptsize eff}}$~~&
~~$\beta P/n$,~$Z=Z_{\hbox{\scriptsize bare}}$~~ & 
~~$Z_{\hbox{\scriptsize eff}}/Z_{\hbox{\scriptsize bare}} $~~ \\
\hline\hline
0.022   &  0.98  &  0.99   &   0.99   &  1.0  \\ \hline
0.044   &  0.95  &  0.96   &   0.97   &  0.99 \\ \hline
0.089   &  0.89  &  0.88   &   0.92   &  0.95 \\ \hline
0.178   &  0.71  &  0.68   &   0.78   &  0.82 \\ \hline
0.356   &  0.45  &  0.41   &   0.46   &  0.53 \\ \hline
0.712   &  0.26  &  0.21   &   -0.32  &  0.29 \\ \hline
\hline
\end{tabular}
\caption{Equation of state (\ref{eq:pressZ}) 
with (third column) and without (fourth column)
renormalization of volume
terms, as a function of electrostatic coupling $\lambda_B/a$, for a 
packing fraction $\eta=0.00125$. The quantity $n$ denotes the mean total
density of counterions $n=\rho_p Z_{\hbox{\scriptsize bare}}$.
The MC data (second column) 
are taken from reference
\cite{Linse}. Since the previous parameters do not correspond to the
saturation regime of effective charges but only approach it, we have
used the effective charge given by Alexander's prescription \cite{Alexander} 
to compute
the pressure from Eq. (\ref{eq:pressZ}) in the third column. The corresponding 
ratio $Z/Z_{\hbox{\scriptsize bare}} $ is indicated in the last column}
\label{table:1}
\end{center}
\end{table}

The above analysis shows that 
different routes to  thermodynamic pressure
lead to very different results.
In order to decide which route is the most reliable, a comparison 
with ``exact results'' is welcome. As a benchmark,
we can use the Monte Carlo (MC) pressure data of Linse \cite{Linse}
for salt-free asymmetric electrolytes consisting of highly charged spherical
macroions and point counterions. 
At high electrostatic 
couplings, this system
exhibits an instability and  separates into two coexisting phases
of different electrolyte concentration. 
We shall argue that a minimum requirement for a reliable
theory of phase behavior is its ability to 
reproduce reasonably accurately the
MC equation, at least up to the transition point.
This appears to be a stringent test and a necessary condition 
to trust any instability that a theory might predict.

We first test in table \ref{table:1} 
the equation of state (\ref{eq:pressZ}) for a charge
asymmetry $Z_{\hbox{\scriptsize bare}}=40$ between colloids and counterions.
It is evident that renormalization of colloidal charge  significantly
improves upon complete neglect of non linearities. The latter approach
consists in considering $Z= Z_{\hbox{\scriptsize bare}}$ and severely fails 
at high $\lambda_B/a$, giving negative pressures. Our renormalized volume term
captures the main effect of nonlinearities, but the agreement with MC, even
if decent in view of the simplicity of the approach is nevertheless
only qualitative, and does not reach the level of accuracy required
to discuss phase stability.

\begin{figure}
\includegraphics[width=8cm]{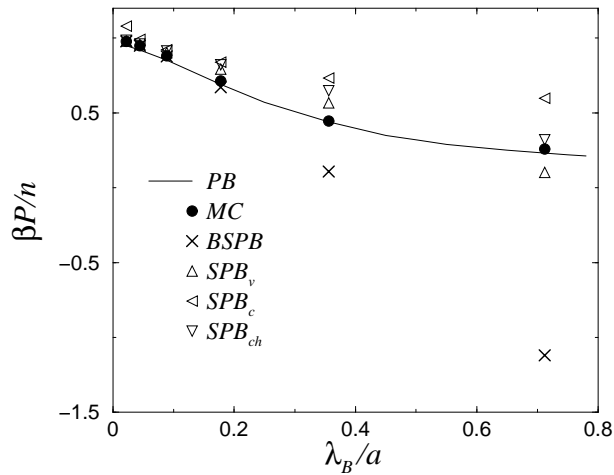}
\caption{Monte Carlo pressures (dots) compared to those obtained within 
PB cell model (continuous curve), boot strap PB \cite{BSPB} (crosses) and
symmetric PB theories \cite{Bhuiyan} 
(triangles, corresponding to virial, charging and compressibility routes).
The packing fraction is $\eta=0.00125$ and $n = \rho_p Z_{\hbox{\scriptsize bare}}$.
As in table \ref{table:1}, the charge asymmetry polyion/counterion is
$Z_{\hbox{\scriptsize bare}}=40$.  }
\label{fig:pb1}
\end{figure}

\begin{figure}
\includegraphics[width=8cm]{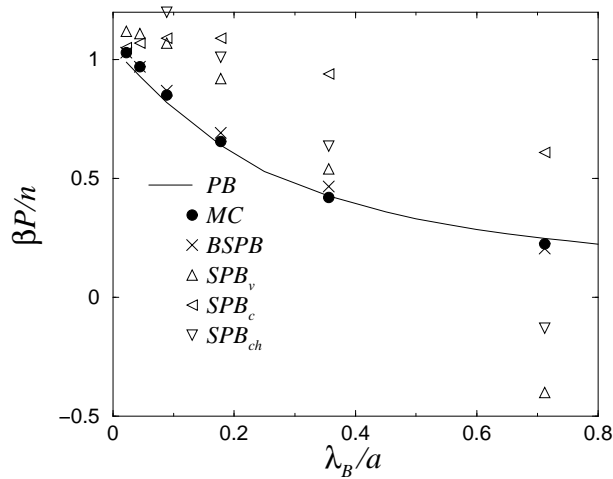}
\caption{Same as Figure \ref{fig:pb1}, for a packing fraction $\eta=0.08$. }
\label{fig:pb2}
\end{figure}

The jellium equation of state derived in section \ref{sec:jellium} only holds 
for saturated effective charges, i.e. in a regime of coupling
that the Monte Carlo simulations, so far, have not reached
(which corresponds to a very high bare charge with
a large separation of scales between Bjerrum length and colloid radius,
see below).  We therefore directly turn to the comparison of
the relative performances of the Poisson-Boltzmann cell model,
symmetric PB \cite{Bhuiyan} and boot-strap PB (see \cite{BSPB} for details),
with respect to Monte Carlo data (see Figures \ref{fig:pb1} and \ref{fig:pb2}).

To produce these figures (providing a similar comparison as Table 
\ref{table:1}), we have chosen the lowest and the highest packing fractions 
investigated by Linse in \cite{Linse}.
The striking feature revealed by Figs \ref{fig:pb1} and \ref{fig:pb2}
is the remarkably good agreement between the PB cell pressures \cite{Wenner,Holm}
and the MC simulations,
even at $\eta=0.00125$ where the cell model could have been anticipated
to fail (see also \cite{Lobaskin}). 
The only competitive approach at $\eta = 0.08$ seems to be the boot strap 
PB theory \cite{BSPB}, but this theory severely fails for low volume
fractions, see Fig \ref{fig:pb1}, necessary to study colloidal 
phase-stability.
At this volume fraction, the simple treatment of section \ref{sec:theory}
[Eq. (\ref{eq:pressZ})], provides a better equation of state 
than the boot strap PB or symmetric PB (see Table \ref{table:1} and Fig. \ref{fig:pb1}).  
Within the PB cell, polyion-polyion as well as counterion-counterion
correlations are discarded~; Figs \ref{fig:pb1} and \ref{fig:pb2} show that 
as far as the pressure is concerned,
these contributions are small or negligible, for the parameters investigated, 
even at the highest couplings. This justifies their neglect in our 
analytical treatment. 

\begin{figure}
\includegraphics[width=8cm]{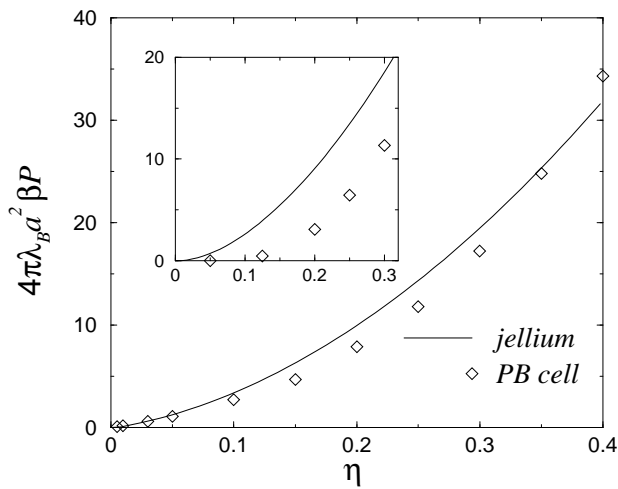}
\caption{Comparison between PB cell pressures and those for the jellium
model, without added salt [in the latter situation, the equation of state is
given by expression (\protect\ref{eq:pressjel})]. 
Inset: same when the suspension is dialyzed against
a salt reservoir such that $\kappa_s a = 2.6$ (the quantity $P$ considered is
the osmotic pressure, i.e. the reservoir contribution has been subtracted).
}
\label{fig:jellium}
\end{figure}

Given the accuracy of the PB cell model, we are now in the position 
to assess the quality of the jellium  approximation 
of section \ref{sec:jellium}. The corresponding pressures are 
compared in Fig. \ref{fig:jellium} with their PB cell counterparts 
for highly charged
colloids (where the effective charge saturates to its upper threshold),
both with and without added salt. The simple analytical expression 
(\ref{eq:pressjel}) for salt free suspensions is found to be 
in good agreement with the PB data. Unfortunately the agreement 
deteriorates when $\kappa_s a>1$ (see the inset).

%%%%%%%%%%%%%%%%%%%%%%%%%%%%%%
\subsection{Relevance for colloidal suspensions}

At this point, we must  conclude that the PB theory, 
even restricted to the cell,
is superior to the competing approaches {\it for aqueous suspensions
with monovalent counterions}.   
At high electrostatic couplings, corresponding to multivalent
counterions in water, the MC simulations of Linse \cite{Linse} find an
instability.  This transition, however, has nothing to do
with the volume terms, but is the result of 
strong correlations  
between the double layers of colloidal particles\cite{Levin}
which produce  attraction between like-charged colloids
at sufficiently short separations\cite{Rouzina,Levin2}.
So far, this attraction has not been properly included in any of the
thermodynamic theories of colloidal stability.

{\em Validity of PB theory}.
To quantify the range of validity of the PB theory one 
may construct a dimensionless
parameter $\Gamma_{cc}$ characterizing 
the importance of microions correlations,
discarded within the PB theory. For monovalent microions, 
$\Gamma_{cc} \propto \beta q^2/(\ell \epsilon)$ where $\ell$ is the 
characteristic
mean distance between the microions in the double layer. 
If the number of condensed counterions is such
as to almost completely neutralize the colloidal charge, which
is the case for strongly charged colloids, 
$\ell \simeq a / \sqrt{Z_{\hbox{\scriptsize bare}}}$,
and  $\Gamma_{cc}$ becomes\cite{Rouzina,Levin}
\begin{equation}
\Gamma_{cc} = \frac{\lambda_B}{a} \sqrt{\frac{Z_{\hbox{\scriptsize bare}}}{4\pi}}.
\end{equation}
When $\Gamma_{cc}$ exceeds unity, PB theory is expected to break down. 
The value $\Gamma_{cc} \simeq 2$ has been reported to correspond to the
instability threshold \cite{Rouzina,Linse_corro}, which and has been
observed in the simulations of Linse \cite{Linse}. 
The field theoretic treatment
of Netz 
%in the planar geometry (where $a$ diverges at constant 
%$Z_{\hbox{\scriptsize bare}}/a^2$) 
also
corroborates this conclusion \cite{Netz}.
For particles with $Z_{\hbox{\scriptsize bare}}=40$, $\Gamma_{cc} \simeq 2$ corresponds
to $\lambda_B/a \simeq 1.1$. Thus the PB theory can be expected
to work quite well up to very high surface charge concentrations. 
Indeed, comparing the predictions of the PB cell model to the
MC simulations, an excellent agreement is observed up to 
$\lambda_B/a \simeq 0.7$ where the MC data stop, rather close to the
expected point of instability $\lambda_B/a \simeq 1.1$, 
see Figs.  \ref{fig:pb1} and \ref{fig:pb2}. 

{\em Validity of the saturation picture within the PB}.
The constant potential approach used in sections \ref{sec:theory}
and \ref{sec:jellium} relies on the phenomenon of effective charge
saturation exhibited by the PB theory, when $Z_{\hbox{\scriptsize bare}}$ 
is large enough. The saturation occurs when the electrostatic
energy of the condensed counterion is significantly larger than
$k_B T$.  This can be characterized by a dimensionless parameter
\begin{equation}
\Gamma_{\hbox{\scriptsize sat}} =\frac{Z_{\hbox{\scriptsize bare}}}{Z_{eff}}=\frac{Z_{\hbox{\scriptsize bare}} \lambda_B}
{4 a (1+\kappa_s a)}.
\end{equation}
When $\Gamma_{\hbox{\scriptsize sat}}$ becomes larger 
than 1 \cite{JPA,Trizac2}, linearized theory fails
and charge renormalization becomes important. 

\begin{figure}
\includegraphics[width=8cm]{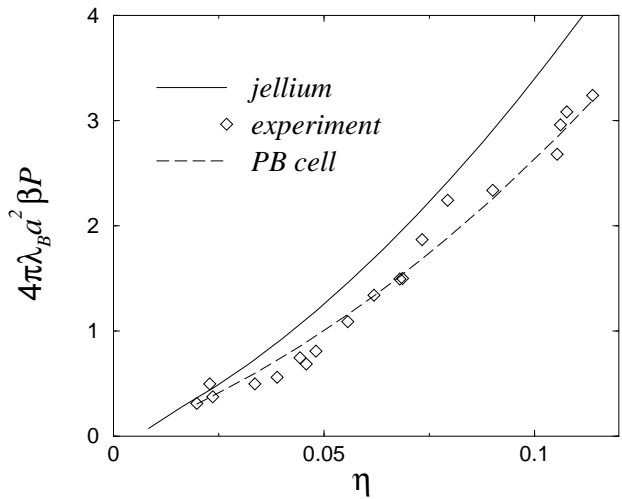}
\caption{Same as figure \ref{fig:jellium}, including a comparison with
osmotic pressures of deionized bromopolystyrene particles (shown by diamonds,
from ref. \cite{Reus}). The curve labeled ``jellium'' corresponds 
to the analytical prediction (\ref{eq:pressjel}). No adjustable parameters have been
used.}
\label{fig:reus}
\end{figure}

We must stress that large values of  $\Gamma_{\hbox{\scriptsize sat}}$
are fully compatible with small values of of $\Gamma_{cc}$.  
Specifically, for any $a \gg \lambda_B$, there exists a range
of bare colloidal charges $Z_{\hbox{\scriptsize bare}}$, such that 
$\Gamma_{cc}<2<\Gamma_{\hbox{\scriptsize sat}}$.  For these
values of $Z_{\hbox{\scriptsize bare}}$, 
the polyion-microion interaction is sufficiently 
strong that linearized theories, without charge renormalization,
will certainly fail.  On the other hand the counterion-counterion
correlations are sufficiently weak, so that the PB theory 
is still applicable. To illustrate this point, we compare in Fig. 
\ref{fig:reus} the pressure obtained at saturation within the PB cell
model calculation
(formally $Z_{\hbox{\scriptsize bare}} \to \infty$), to that
measured experimentally by Reus {\it et al}, under the conditions
very close to complete deionization (no salt). The agreement with
PB theory had already been mentioned in \cite{Reus,Trizac2}, but provides
an illustration of the saturation phenomenon in real suspensions.
It also shows that despite its simplicity, Eq. (\ref{eq:pressjel}) is fairly accurate.

It is important to stress, however, that 
for any colloidal size $a$, there is a maximum
value of $Z_{\hbox{\scriptsize bare}}$ above which the  PB theory
fails. In practice, however, this break down of the
PB equation never occurs for aqueous 
suspensions containing only monovalent counterions.
The reason for this is that the hydrated ionic size provides
a lower cutoff for the length $\ell$ which appears in $\Gamma_{cc}$, i.e.
$\ell>d$, where $d\approx 4$ \AA, is the hydrated ionic diameter.
For aqueous suspension with monovalent counterions $\Gamma_{cc}$
is, therefore, always less than $\lambda_B/d$, so that  
$\Gamma_{cc}<2$. For monovalent counterions in water,
the PB theory, therefore, should apply without any restriction.  Of course,
this pleasant situation changes as soon as  
multivalent salt is added to suspension.
In this case,  hydrated ionic size is no longer sufficient
to restrict the strength of microion-microion correlations and 
$\Gamma_{cc}>2$. Under these conditions the PB theory will no
longer apply and a more sophisticated approach must be used.

%%%%%%%%%%%%%%%%%%%%%%%%%%%%%%%%%%%%%%%%%%%%%%%%%%%%%%%%%%%%%%%%%%%%%%%%
\section{Summary and conclusions}
\label{sec:conclusion}

We have proposed (section \ref{sec:theory}) a linear theory
to investigate the phase behavior of colloidal suspensions. 
The  non-linear effects are partially taken into account 
through
the postulate that  highly charged polyions
behave as if they were  constant potential objects.
The effective charge of colloidal particles is, therefore,
a state dependent function.
Our first goal was to develop a consistent thermodynamic approach 
for such state dependent charges. The results found in section \ref{sec:theory}
rely on a simple form for the electrostatic free energy (volume term), 
resulting from the polyion-microion interactions calculated using the
linearized PB equation, for spherical colloids. 
The approach could be easily generalized to the case of cylindrical macroions.
The critical behavior
predicted is, however, 
spurious, which may be attributed to the simplicity of the volume term used, 
and/or the difficulty of renormalizing such terms.
The jellium-like model of section \ref{sec:jellium} provides 
a more reliable route, and allows to obtain analytically a simple equation of
state for highly charged colloids in the salt free limit, see 
Fig. \ref{fig:reus}. Unfortunately it is difficulty to see how
this kind of approach can be extended to account for 
the polyion-polyion interactions.

>From our analysis, we conclude that for 
$\Gamma_{cc} = [Z_{\hbox{\scriptsize bare}}/(4 \pi)]^{1/2} \lambda_B/a < 2 $,
Poisson-Boltzmann approach (PB), even restricted to the cell model,
leads to more accurate predictions for the
thermodynamic functions than the  
competing theories. This is quite remarkable, since it is
by far the simplest (see the appendix of 
\cite{Langmuir} for a ``ready-to-use'' implementation
of PB cell model). However, there is clearly a need to go beyond the PB
theory when dealing with the multivalent counterions, since it is the 
counterion-counterion
correlations   that drive a phase instability 
for $\Gamma_{cc}>2$. Inclusion of these effects 
in a theoretical approach is a difficult task, since they have 
little signature on the pressure data up to the electrostatic coupling where 
suddenly they destabilize the system.

%Finally, as far as the real aqueous  suspensions 
%with monovalent counterions are concerned, the inequality
%$\Gamma_{cc} <2$ is systematically obeyed
%which indicates that the PB theory should be valid and that suspensions
%should be stable against fluid-fluid phase separation. The 
%phase instability  predicted for these systems by linear
%theories is apparently an artifact of the underlying approximations.
% 
%For the 
%validity of our starting point in sections \ref{sec:theory} and \ref{sec:jellium},
%we have also justified 
%the physical and experimental relevance of the saturation
%of effectives charges observed within PB at intermediate electrostatic
%coupling. This point is sometimes confused in the literature, and appears
%to contradict a conclusion put forward by Onsager in a seminal paper \cite{Onsager}
%\begin{quote}
%{\em As soon as the higher order [{\rm non-linear}] terms in the Poisson-Boltzmann
%equation become important, we can no longer expect the ionic atmosphere 
%to be additive, and then the Poisson-Boltzmann equation itself becomes
%unreliable.}
%\end{quote} 
%This statement (irrelevance of PB as soon as non linear behavior sets in)
%is justified for simple electrolytes (the system studied by Onsager in \cite{Onsager}),
%but not for colloids.

This work was supported in part by the Brazilian agencies
%Conselho Nacional de Desenvolvimento Cient{\'\i}fico e Tecnol{\'o}gico 
CNPq and FAPERGS and by the french CNRS. Y. Levin acknowledges the
hospitality of the Theoretical Physics Laboratory in Orsay where
part of this work was performed.

%%%%%%%%%%%%%%%%%%%%%%%%%%%%%%%%%%%%%%%%%%%%%%%%%%%%%%%%%%%%%%%%%%%                                   
%\bibliographystyle{prsty}
%\bibliography{references}

\end{document}